\documentclass[aps,preprint,showpacs,groupedaddress,floatfix]{revtex4}%
\usepackage{graphicx}
\usepackage{dcolumn}
\usepackage{bm}
\usepackage{amsmath}
\usepackage{amsfonts}
\usepackage{amssymb}%
\newcommand{\be}{\begin{equation}}
\newcommand{\ee}{\end{equation}}
\newcommand{\bea}{\begin{eqnarray}}
\newcommand{\eea}{\end{eqnarray}}

\begin{document}
\title{%
Covariant response theory beyond RPA and its application}
\author{\firstname{E.}~\surname{Litvinova}}
\email{elena.litvinova@ph.tum.de}
\affiliation{Physik-Department der Technischen Universit\"at
M\"unchen, D-85748 Garching, Germany}
\affiliation{Institute of
Physics and Power Engineering, 249020 Obninsk, Russia}
\author{\firstname{P.}~\surname{Ring}}
\email{peter.ring@ph.tum.de}
\affiliation{Physik-Department der Technischen Universit\"at M\"unchen,
D-85748 Garching, Germany}
\author{\firstname{V.}~\surname{Tselyaev}}
\email{tselyaev@nuclpc1.phys.spbu.ru}
\affiliation{Nuclear Physics Department,
 V. A. Fock Institute of Physics,
 St. Petersburg State University, 198504,
 St. Petersburg, Russia}
%
%
\date{\today}
\begin{abstract}
The covariant particle-vibration coupling model within the time blocking approximation is
employed to supplement the Relativistic Random Phase Approximation (RRPA) with
coupling to collective vibrations. The Bethe-Salpeter equation
in the particle-hole channel with an energy dependent residual particle-hole (p-h)
interaction is formulated and solved in the shell-model Dirac basis as well
as in the momentum space. The same set of the coupling
constants generates the Dirac-Hartree single-particle spectrum, the static part of the
residual p-h interaction and the particle-phonon coupling amplitudes.
This approach is applied to quantitative description of damping phenomenon
in even-even spherical nuclei with closed shells $^{208}$Pb and $^{132}$Sn.
Since the phonon coupling enriches the RRPA spectrum with a multitude of
ph$\otimes$phonon states a noticeable fragmentation of giant monopole
and dipole resonances is obtained
in the examined nuclei. The results are compared with experimental data and with
results of the non-relativistic approach.

\end{abstract}

\pacs{21.10.-k, 21.60.-n, 24.10.Cn, 21.30.Fe, 21.60.Jz, 24.30.Gz}
\maketitle

\section{Introduction}

Recent development of experimental facilities with radioactive
nuclear beams has stimulated enhanced efforts on the theoretical
side to understand the dynamics of the nuclear many-body problem by
microscopic methods. The most successful schemes based on the mean
field concept use a phenomenological ansatz incorporating as many
symmetries of the system as possible and adjust the parameters of
functionals to ground state properties of characteristic nuclei all
over the periodic table. Of particular interest are the models based
on covariant density functionals \cite{Rin.96,VALR.05} because of
their Lorentz invariance. A large variety of nuclear phenomena have
been described over the years within this kind of models: the
equation of state in symmetric nuclear matter, ground state
properties of finite spherical and deformed nuclei all over the
periodic table \cite{GRT.90} from light nuclei \cite{LVR.04a} to
super-heavy elements \cite{LSRG.96}, from the neutron drip line,
where halo phenomena are observed \cite{MR.96} to the proton drip
line \cite{LVR.04} with nuclei unstable against the emission of
protons \cite{LVR.99}. In the small amplitude limit one obtains the
relativistic Random Phase Approximation (RRPA) \cite{RMG.01}. This
method provides a natural framework to investigate collective and
non-collective excitations of $ph$-character. It is successful in
particular for the understanding of the position of giant resonances
and spin- or/and isospin-excitations as the Gamov Teller Resonance
(GTR) or the Isobaric Analog Resonance (IAR). Recently it has been
also used for a theoretical interpretation of low-lying dipole
\cite{PRN.03} and quadrupole \cite{Ans.05} excitations.

Of course the density functional theory based on the mean field
framework cannot provide an exact treatment of the full nuclear
dynamics. It is known to break down  already in ideal shell-model
nuclei such as $^{208}$Pb with closed protons and neutron shells. In
self-consistent mean field calculations one finds usually a
considerably reduced level density at the Fermi surface as compared
with the experiment. The RRPA describes very well positions of giant
resonances but underestimates their width considerably. To solve the
level density problem the covariant theory of particle-vibration
coupling has been developed and applied in the Ref. \cite{LR.06}. In
the present work we formulate the covariant response theory employing
the particle-vibration coupling model within the time-blocking
approximation \cite{Ts.89,KTT.97,Ts.05,LT.05} to describe the
spreading of multipole giant resonances in even-even spherical
nuclei.
\section{Formalism}
In the relativistic nuclear mean-field
theory motion of a single nucleon is
described by the Dirac equation with an effective mass $m^{\ast}$
and a generalized four-vector of the momentum $P^{\mu}$:
\begin{equation}
\bigl(\gamma^{\mu}P_{\mu}-m^{\ast}\bigr)|\psi\rangle=0\,.
\label{emass}%
\end{equation}
These quantities are determined
by the scalar $\tilde{\Sigma}_{s}$
and vector
$\tilde{\Sigma}^{\mu}=
(\tilde{\Sigma}^{0},\mathbf{\tilde{\Sigma}})$
parts of the mass operator (self-energy) $\tilde{\Sigma}$
within mean-field approximation:
\begin{equation}
m^{\ast}=m+\tilde{\Sigma}_{s}\,,\qquad
P_{\mu}=p_{\mu}-\tilde{\Sigma}_{\mu}=
\Bigl(i\frac{\partial}{\partial t}-\tilde{\Sigma}_{0}
,i\mathbf{\nabla}+\mathbf{\tilde{\Sigma}}\Bigr)\,,
\end{equation}
$\tilde{\Sigma}_{s}$ is generated by the scalar $\sigma$-meson field.
When we go beyond the mean-field approximation,
we have to take into account that
in the general case the full self-energy
$\Sigma$ is non-local in the space
coordinates and also in time. This non-locality means that its
Fourier transform has both momentum and energy dependence.
Let us decompose the total self-energy
matrix into two components,
a static local and and an energy dependent non-local term:
\begin{equation}
\Sigma(\mathbf{r},\mathbf{r^{\prime}};\varepsilon)=
{\tilde{\Sigma}}(\mathbf{r}%
)\delta(\mathbf{r}-\mathbf{r^{\prime}})+\Sigma^{e}(\mathbf{r}%
,\mathbf{r^{\prime}};\varepsilon),
\label{sedecom}
\end{equation}
where index "e" indicates the energy dependence. Due to
time-reversal symmetry and the absence of currents the space-like
components of $\tilde{\Sigma}$ vanish, therefore only scalar and the
time-like components of the mean field are considered in the following.

The quantity ${\tilde\Sigma}({\bf r})$ is assumed to be the RMF
self-energy generated by the $\sigma$, $\omega$ and $\rho$-meson
fields within the framework of the no-sea approximation (see, for
instance, Ref \cite{RMG.01}). To describe the non-local part
$\Sigma^{e}(\mathbf{r},\mathbf{r^{\prime}};\varepsilon)$ we apply the
covariant version of the particle-phonon coupling model
\cite{LR.06}. Due to  the decomposition (\ref{sedecom}) it is
convenient to work in the shell-model Dirac basis
$\{|\psi_{k}\rangle\}$ which diagonalizes the energy-independent
part of the Dirac equation:
\begin{equation}
h^{\mathcal{D}}|\psi_{k}\rangle=\varepsilon_{k}|\psi_{k}\rangle,
\ \ \ \ \ \
h^{\mathcal{D}}={\mbox{\boldmath $\alpha$}}\mathbf{p}+\beta(m+{\tilde{\Sigma}%
}_{s})+{\tilde{\Sigma}}_{0},
\end{equation}
where $h^{\mathcal{D}}$ denotes the Dirac hamiltonian with the
energy-independent mean field. In the case of spherical simmetry the
spinor $|\psi_k\rangle$ is characterized by the set of
single-particle quantum numbers $k = \{(k), m_k \}, (k) = \{n_k,
j_k,\pi_k, t_k \}$ with the radial quantum number $n_k$, angular
momentum quantum numbers $j_{k},m_{k}$, parity $\pi_{k}$ and isospin
$t_{k}$. In this basis the Dyson equation for the single-particle
Green's function can be formulated as follows:
\begin{equation}
\sum\limits_{l}\bigl\{(\varepsilon-\varepsilon_{k})\delta_{kl}-\Sigma_{kl}%
^{e}(\varepsilon)\bigr\}G_{lk^{\prime}}(\varepsilon)=\delta_{kk^{\prime}}.
\label{fg1}%
\end{equation}
Matrix elements of the energy-dependent part of the mass operator
%
%
%
\begin{equation}
\Sigma_{kl}^{e}(\varepsilon)=\int d^{3}rd^{3}{r}^{\prime}~
\bar{\psi}_{k}(\mbox{\boldmath $r$})\Sigma^{e}%
({\mbox{\boldmath $r$}},{\mbox{\boldmath $r$}^{\prime}};\varepsilon)
\psi_{l}({\mbox{\boldmath $r$}^{\prime}})
\label{Sigmae}%
\end{equation}
are expressed in terms of the particle-phonon coupling model
\begin{equation}
\Sigma_{kl}^{e}(\varepsilon)= \sum\limits_{q,n}
\frac{\gamma_{kn}^{q(\sigma_{n})}\,
\gamma_{ln}^{q(\sigma_{n})\ast}}
{\varepsilon-\varepsilon_{n}-\sigma_{n}(\Omega^{q}-i\eta)},
\qquad
\gamma_{kn}^{q(\sigma)} =
\delta^{\vphantom{A}}_{\sigma,+1}\gamma_{kn}^{q} +
\delta^{\vphantom{A}}_{\sigma,-1}\gamma_{nk}^{q\ast}
\label{mo}%
\end{equation}
through the phonon vertexes $\gamma^{q}$ and their frequencies
$\Omega^{q}$. They are determined by the following relation:
\begin{equation}
\gamma_{kl}^{q}=\sum\limits_{k^{\prime}l^{\prime}}
V^{\vphantom{A}}_{kl^{\prime},lk^{\prime}}
{\hat\rho}^{q}_{k^{\prime}l^{\prime}},
\qquad
V_{kl^{\prime},lk^{\prime}} =
\frac{\delta{\tilde\Sigma}_{l^{\prime}k^{\prime}}}
{\delta\rho_{lk}}.
\label{phonon}%
\end{equation}
$V_{kl^{\prime},lk^{\prime}}$ denotes the matrix element of the
residual interaction which is a functional derivative of the
relativistic mean field with respect to nuclear density $\rho$ and
${\hat\rho}^q$ is the transition density. Here we use the
linearized version of the model which assumes that ${\hat\rho}^q$
is not influenced by the particle-phonon coupling and can be computed
within the relativistic RPA. In the present work the residual
interaction is generated by the relativistic NL3 Lagrangian
\cite{NL3}. In the Eq. (\ref{mo}) $\sigma_{n}=+1$ if $n$ is an
unoccupied state of $p$- or $\alpha$-types and $\sigma_{n}=-1$ for
an occupied $n$ state of $h$-type, $\eta \to +0$. $\alpha$ denotes
states in the Dirac sea with negative energies which arise in the
Lehmann expansion of the single-particle Green's function due to the
no-sea approximation.

Equation (\ref{fg1}) has been solved numerically in the Ref.
\cite{LR.06} in the shell-model of Dirac states. A noticeable
increase of the single-particle level density near the Fermi surface
relative to the pure RMF spectrum is obtained for $^{208}$Pb. This
improves the agreement of the single-particle level scheme with
experimental data considerably. For the four odd mass nuclei
surrounding $^{208}$Pb the distribution of the single-particle
strength has been calculated and compared with experiment as well as
with the results obtained within several non-relativistic
approaches.

\begin{figure}[ptb]
\begin{center}
\includegraphics*[scale=0.75]{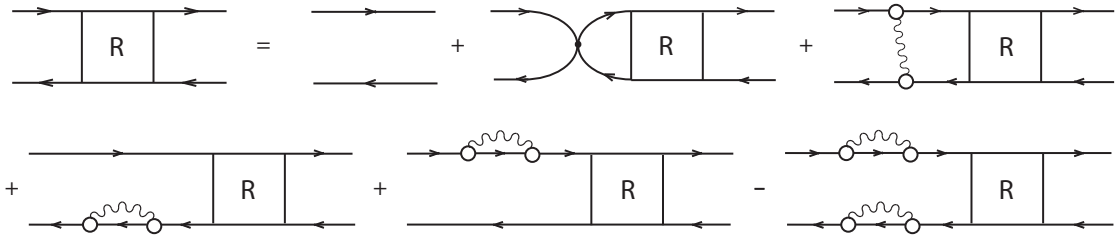}
\end{center}
\caption{Bethe-Salpeter equation for the p-h response function in
the graphical representation. Solid lines with arrows denote
one-body propagators through the particle, hole or antiparticle
states, weavy lines denote phonon propagators, empty circles are the
particle-phonon coupling amplitudes and the small black circle means
the static part of the residual p-h interaction.
}%
\label{f1}%
\end{figure}

The nuclear dynamics of an even-even nucleus in a weak external
field is described by the linear response function which is a
solution of the Bethe-Salpeter equation (BSE) in the particle-hole
(p-h) channel. In the beginning it is convenient to consider this
equation in the time representation. Let us include the time
variable into the set of single-particle quantum numbers and use the
following number indexation to simplify the expressions: $1 =
\{k_1,t_1\}$. In this notation the BSE for the response function $R$
reads:
\begin{equation}
R(14,23) = -{\tilde G}(1,3){\tilde G}(4,2) +
\frac{1}{i}\sum\limits_{5678}
{\tilde G}(1,5){\tilde G}(6,2)W(58,67)R(74,83), \label{bse1}
\end{equation}
where
\begin{equation}
W(14,23) = U(14,23) + i\Sigma^e(1,3){\tilde G}^{-1}(4,2) +
i{\tilde G}^{-1}(1,3)\Sigma^e(4,2) -
i\Sigma^e(1,3)\Sigma^e(4,2).
\label{wampl}
\end{equation}
Here the summation over number indices implies also integration over
respective time variables. $\tilde G$ is the mean field
single-particle Green's function
%
%
and $U$ is irreducible in the p-h channel amplitude
of the effective interaction
including an induced interaction due to the phonon exchange.
The graphical representation of the Eq. (\ref{bse1}) is shown in
Fig. \ref{f1}.

Using Fourier transformation of the Eq. (\ref{bse1}) one comes to an
integral equation where both the solution and the kernel are
singular with respect to energy variables. Another difficulty
arises because the Eq. (\ref{bse1}) contains integrations over all
time points of the intermediate states. This means that many
configurations which are actually more complex than
1p1h$\otimes$phonon are contained in the exact response function. In
the Ref. \cite{Ts.89} the special time-projection technique was
introduced to block the p-h propagation through these complicated
intermediate states. It has been shown that for this type of
response it is possible to reduce the integral equation to a
relatively simple algebraic equation. Obviously, this method can be
applied straightforwardly to our case. The full formalism can be
found in the Ref. \cite{RLT.06}. Making use the above mentioned time
projection one can
transform the Eq. (\ref{bse1}) to the following
algebraic equation within
the so-called time-blocking approximation:
\bea
R_{k_1k_4,k_2k_3} (\omega) &=& \tilde{R}_{k_1k_4,k_2k_3} (\omega)
\nonumber\\
&-& \sum\limits_{k_5k_6k_7k_8}
\tilde{R}_{k_1k_6,k_2k_5} (\omega)\,
\bigl[  V_{k_5k_8,k_6k_7} + \Phi_{k_5k_8,k_6k_7} (\omega)\,\bigr]\,
R_{k_7k_4,k_8k_3} (\omega)\,,
\label{bse2}
\eea
where $\tilde{R}$ is the mean-field p-h propagator,
$V$ is the residual interaction defined by the Eq.~(\ref{phonon}),
$\Phi$ is the particle-phonon coupling amplitude
including phonon contribution both into the self-energy
and into the induced interaction. It is supposed that
the summation is carried out over the whole Dirac space.

Thus, to describe the observed spectrum of the excited nucleus
in the weak external field $P$
within this formalism one needs to solve the Eq.~(\ref{bse2}) and
to calculate the strength function:
\begin{equation}
S(E) = \frac{1}{\pi}\lim\limits_{\Delta\to +0}\,
\mbox{Im}\sum\limits_{k_1k_1k_3k_4}
P_{k_1k_2}^{\ast}R_{k_1k_4,k_2k_3}(E+i\Delta)P_{k_3k_4}.
\end{equation}
The imaginary part $\Delta$ of the energy variable is introduced for convenience
in order to obtain more smoothed envelope of the spectrum. This parameter has a meaning of an
additional artificial width for each excitation. This width emulates 
effectively contributions from configurations which are not taken into account 
explicitely.
\section{Results and discussion}
The developed approach is applied to a quantitative description of
isoscalar monopole and isovector dipole giant resonances in the
even-even spherical nuclei $^{208}$Pb and $^{132}$Sn. Details of our
calculation scheme are given in the Ref. \cite{RLT.06}. First, the
Dirac equation for single nucleons together with the Klein-Gordon
equations for meson fields (RMF problem) are solved simultaneously
to obtain the single-particle basis. Second, the RRPA equations
\cite{RMG.01} are solved to determine for the above mentioned
phonons. These two sets form the multitude of ph$\otimes$phonon
configurations which enter the particle-phonon coupling amplitude
$\Phi$. Third, an equation for density matrix variation 
(convolution of the Eq. (\ref{bse2}) with the external field
operator) is solved with this additional amplitude. It provides an
enrichment of the calculated spectrum as compared to the pure RRPA.
The equation for density matrix variation has been solved both
in the momentum and in the Dirac spaces to ensure propriety of our
calculational scheme and identical results have been obtained.
\begin{figure}[ptb]
\begin{center}
\includegraphics*[scale=1.2]{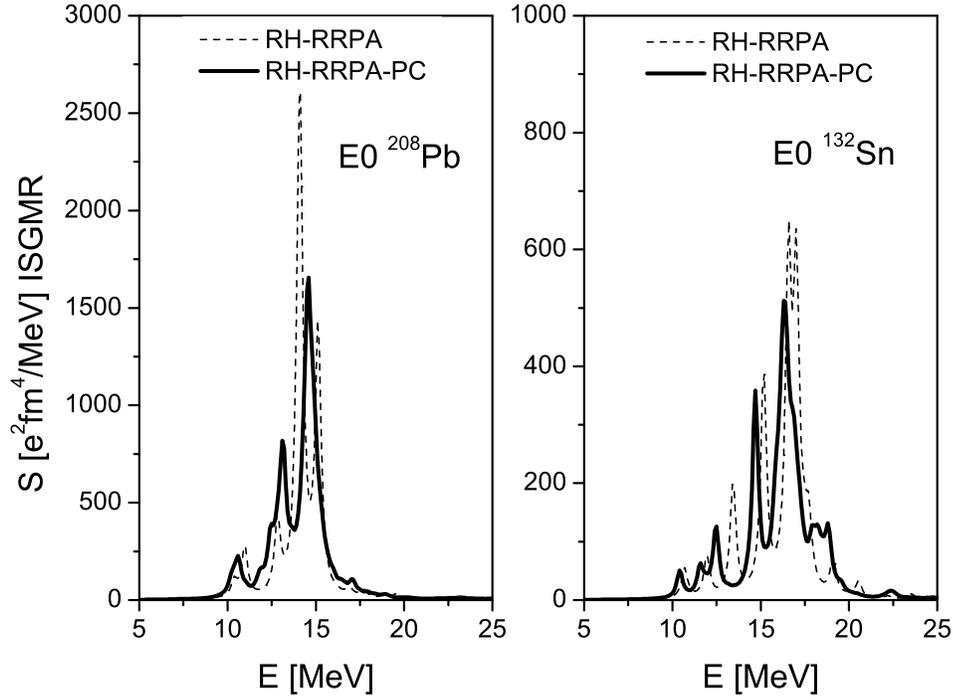}
\end{center}
\caption{Isocalar monopole resonance in $^{208}$Pb and $^{132}$Sn
obtained within two approaches: the RRPA (dashed lines) and the RRPA with
the particle-phonon coupling RRPA-PC (solid lines). Both calculations
are based on the relativistic Hartree (RH) approach
with the parameter set NL3.}%
\label{f2}%
\end{figure}
\begin{table}[ptb]
\caption{Lorentz fit parameters of isoscalar E0 strength function
in $^{208}$Pb and $^{132}$Sn} calculated within the RRPA and the RRPA extended by
the particle-phonon coupling model (RRPA-PC) as compared to experimental data.%
\label{tab1}
\begin{center}
\vspace{3mm} \tabcolsep=2.15em \renewcommand{\arraystretch}{1.1}%
\begin{tabular}
[c]{cccc}
\hline\hline
 &  & $<$E$>$ (MeV) & $\Gamma$ (MeV) 
\\\hline
 & RRPA & 14.13 & 1.17 
\\
$^{208}$Pb & RRPA-PC & 14.02 & 1.57 
\\
 & Exp. \cite{SY.93}&  13.73(20) & 2.58(20) 
\\
\hline
 & RRPA & 16.13 & 1.96 
\\
$^{132}$Sn & RRPA-PC & 16.07 & 2.37 
\\
\hline\hline
\end{tabular}
\end{center}
\end{table}
The calculated strength functions for the isoscalar monopole
resonance in $^{208}$Pb and $^{132}$Sn computed within the RRPA and the RRPA
extended by the particle-vibration coupling (RRPA-PC) are given in
the Fig. \ref{f2}. The fragmentation of the resonance caused by the
particle-phonon coupling is clearly demonstrated although the spreading
width of the monopole resonance is not large 
because of a strong cancellation between the self-energy diagrams and
diagrams with the phonon exchange (see Fig. \ref{f1}). The mean energies
and widths of these resonances are presented in the Table
\ref{tab1}. As experimental data we display the numbers adopted in
the Ref. \cite{SY.93} for the calculation of the nuclear matter
compressibility from the evaluation of a series of data obtained in
different experiments for the isoscalar monopole resonance in
$^{208}$Pb.
\begin{figure}[ptb]
\begin{center}
\includegraphics*[scale=1.2]{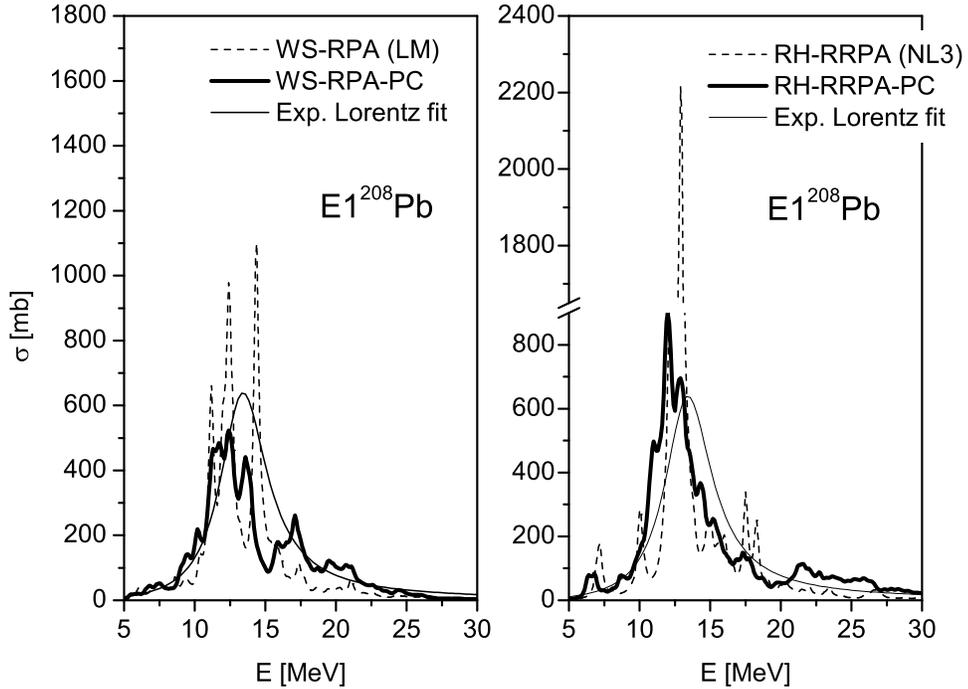}
\end{center}
\caption{Isovector E1 resonance in $^{208}$Pb: the results obtained within
the non-relativistic approach (left panel) with Woods-Saxon (WS) single-particle input
and Landau-Migdal (LM) forces and calculations performed within the covariant
theory (right panel) based on the relativistic Hartree (RH) approach
with the NL3 mean field parameter set.
The RPA calculations are shown by the dashed curves, the RPA-PC calculations 
-- by the thick solid curves.
Experimental Lorentzian is given by the thin solid curves.
}%
\label{f3}%
\end{figure}
\begin{figure}[ptb]
\begin{center}
\includegraphics*[scale=1.2]{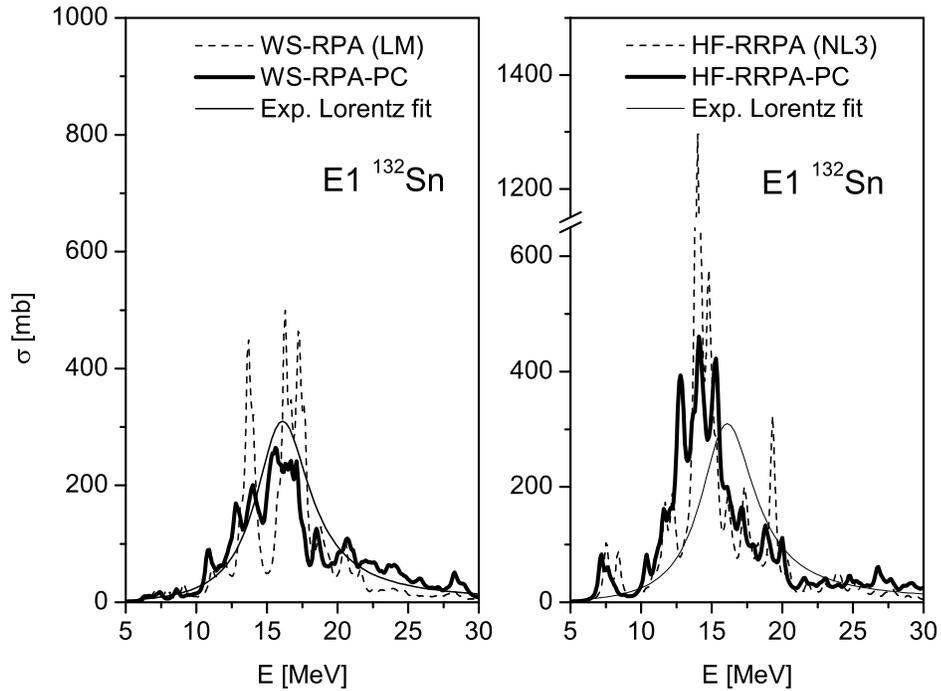}
\end{center}
\caption{The same as in Fig. \ref{f3} but for $^{132}$Sn.}%
\label{f4}%
\end{figure}
\begin{table}[ptb]
\caption{Lorentz fit parameters of the E1 photoabsorption
cross section in $^{208}$Pb and $^{132}$Sn} calculated within the RRPA and the RRPA extended by
the particle-phonon coupling model (RRPA-PC) as compared to experimental data.%
\label{tab2}
\begin{center}
\vspace{3mm} \tabcolsep=2.15em \renewcommand{\arraystretch}{1.1}%
\begin{tabular}
[c]{ccccc}
\hline\hline
 &  & $<$E$>$ (MeV) & $\Gamma$ (MeV) & EWSR (\%) 
\\\hline
 & RRPA & 13.1 & 2.5 & 120 
\\
$^{208}$Pb & RRPA-PC & 12.8 & 3.8 & 114
\\
 & Exp. \cite{ripl} & 13.4 & 4.1 & 
\\
\hline
 & RRPA & 14.7 & 3.0 & 112 
\\
$^{132}$Sn & RRPA-PC & 14.3 & 3.8 & 108
\\
 & Exp. \cite{A132} & 16.1(7) & 4.7(2.1) & 
\\\hline\hline
\end{tabular}
\end{center}
\end{table}

The calculated photoabsorption cross sections for the
isovector dipole resonance in $^{208}$Pb and $^{132}$Sn are given in
the Figs. \ref{f3} and \ref{f4} respectively. The left panels show
the results obtained within the non-relativistic
semi-phenomenological 
approach developed in the Ref. \cite{LT.05}
which includes particle-phonon coupling on the base of
Woods-Saxon single-particle input and Landau-Migdal forces. 
The right panels show the results of
the calculations within the relativistic approach developed in the
present work. To make the comparison reasonable, calculations within
the non-relativistic framework have been performed with box boundary
conditions for the Schr\"{o}dinger equation in the coordinate
representation which ensures completeness of the single-particle
basis. In both cases one can see the noticeable fragmentation
of the resonances due to the particle-phonon coupling. Moreover, one
can find more or less the same level of agreement with experimental
data for these two calculations. In case of the isovector E1
resonance in $^{132}$Sn this is, however, not so clear because the
integral characteristics of the resonance obtained in the experiment
of tthe Ref. \cite{A132} are given with relatively large discrepancies.
But the difference which is of fundamental importance is that in the
semi-phenomenological approach one usually fits the parameters on all
three stages of the calculation: first, the Woods-Saxon well depth is
varied to obtain single-particle levels equal to experimental
values, second, one of the Landau-Migdal force parameters is
adjusted to get phonon energies at the experimental positions (for
each mode) and, third, another Landau-Migdal force
parameter is varied to reproduce
the centroid of the giant resonance. Although the varying
of the parameters is performed in relatively narrow limits,
in some cases 
it is necessary to obtain realistic results.
In contrast, within 
the fully covariant microscopic approach developed in the present
work no adjustment of parameters is made.

The mean energies and widths of the isovector E1 resonance computed within
the RRPA and the RRPA-PC are displayed in the Table \ref{tab2}.
For the isovector E1 resonance as well as for the isoscalar E0 resonance Lorentz fits
have been made according to the method developed in the Ref. \cite{TLor} within the energy
region between one and three neutron separation energies.
For the E1 resonance in $^{132}$Sn the lower energy
limit is slightly higher in order to separate the distinct group of pygmy states 
from the giant resonance.
\section{Summary}

The relativistic random phase approximation is extended by the particle-vibrational coupling model.
The
Bethe-Salpeter equation is formulated in the two-body basis of Dirac states.
Amplitude of the effective interaction entering this equation contains the static part
originating
from the pure relativistic mean field as well as the energy-dependent part caused by the
particle-vibrational coupling. The latter term has been considered within the time-blocking approximation
in a fully relativistic way using the covariant form of the nucleon mass operator.

The developed approach is applied to the computation of
spectroscopic characteristics of nuclear excited states in a wide
energy range up to 30 MeV for even-even spherical nuclei. An
equation for the density matrix variation is solved
in the Dirac space as well as in the momentum space. The
particle-phonon coupling amplitudes of collective vibrational modes
below the neutron separation energy have been computed within the
self-consistent RRPA using the parameter set NL3 for the Lagrangian.
The same force has been employed in giant resonance calculations for
the static part of the effective p-h interaction. Therefore a fully
consistent description of giant resonances is performed.

Noticeable fragmentation  of the isoscalar monopole and isovector
dipole giant resonances in $^{208}$Pb and $^{132}$Sn is obtained due
to the particle-vibrational coupling. This leads to the appearance
of a significant spreading width as compared to RRPA calculations.
This is in agreement with experimental data as well as with the
results obtained within the non-relativistic approaches
\cite{LT.05,SBC}.

\begin{acknowledgments}
This work has been supported in part by the Bundesministerium f\"ur Bildung
und Forschung under project 06 MT 193. E. L. acknowledges the support from the
Alexander von Humboldt-Stiftung and the assistance and hospitality provided
by the Physics Department of TU-M\"unchen. V.~T. acknowledges financial support
from the
Deutsche Forschungsgemeinschaft under the grant No. 436 RUS 113/806/0-1
and from the Russian Foundation for Basic Research under the grant
No. 05-02-04005-DFG\_a.
\end{acknowledgments}
\newpage
%

\begin{thebibliography}{99}
%
\bibitem{Rin.96}
P. Ring, Prog. Part. Nucl. Phys. {\bf 37},  193  (1996).
%
\bibitem{VALR.05}
D. Vretenar, A.~V. Afanasjev, G.~A. Lalazissis, and P. Ring, Phys. Rep.
  {\bf 409},  101  (2005).
%
\bibitem{GRT.90}
Y.~K. Gambhir, P. Ring, and A. Thimet, Ann. Phys. (N.Y.) {\bf 198},  132
  (1990).
%
\bibitem{LVR.04a}
G. Lalazissis, D. Vretenar, and P. Ring, Eur. Phys. J. {\bf A22},  37  (2004).
%
\bibitem{LSRG.96}
G.~A. Lalazissis, M.~M. Sharma, P. Ring, and Y.~K. Gambhir, Nucl. Phys. {\bf
  A608},  202  (1996).
%
\bibitem{MR.96}
J. Meng and P. Ring, Phys. Rev. Lett. {\bf 77},  3963  (1996).
%
\bibitem{LVR.04}
G.~A. Lalazissis, D. Vretenar, and P. Ring, Phys. Rev. C {\bf 69},  017301  (2004).
%
\bibitem{LVR.99}
G.~A. Lalazissis, D. Vretenar, and P. Ring, Nucl. Phys. {\bf A650},  133
  (1999).
%
\bibitem{RMG.01}
P. Ring, Z.-Y. Ma, N. Van~Giai, D. Vretenar, A. Wandelt, and L.-G. Cao, Nucl.
  Phys. {\bf A694},  249  (2001).
%
\bibitem{PRN.03}
N. Paar, P. Ring, T. Nik{\v{s}}i{\'{c}}, and D. Vretenar, Phys. Rev. C {\bf 67},
   034312  (2003).
%
\bibitem{Ans.05}
A. Ansari, Phys. Lett. {\bf B623},  37  (2005).
%
\bibitem{LR.06}
E. Litvinova and P. Ring, Phys. Rev. C {\bf 73}, 044328 (2006).
%
\bibitem{Ts.89}
V.I. Tselyaev, Yad. Fiz. {\bf 50}, 1252 (1989) [Sov. J. Nucl. Phys.
{\bf 50}, 780 (1989)].
%
\bibitem{KTT.97}
S.P. Kamerdzhiev, G.Ya. Tertychny, and V.I. Tselyaev,
Phys. Part. Nucl. {\bf 28}, 134 (1997).
%
\bibitem{Ts.05}
V.I. Tselyaev, arXiv:nucl-th/0505031.
%
\bibitem{LT.05} E.V. Litvinova and V.I. Tselyaev, arXiv:nucl-th/0512030.
%
\bibitem{NL3}
G.~A. Lalazissis, J. K{\"{o}}nig, and P. Ring, Phys. Rev. C {\bf 55},  540
  (1997).
%
%
%
%
%
%
%
%
%
\bibitem{RLT.06} P. Ring, E. Litvinova, V. Tselyaev, to be published.
%
\bibitem{SY.93}
S. Shlomo and D.H. Youngblood, Phys. Rev. C {\bf 47}, 529 (1993).
%
\bibitem{ripl}
Reference Input Parameter Library, Version 2, http://www-nds.iaea.org/RIPL-2/.
%
\bibitem{A132}
P. Adrich, A. Klimkiewicz, M. Fallot et al.,
Phys. Rev. Lett. {\bf 95}, 132501 (2005).
%
\bibitem{TLor}
 V. I. Tselyaev,
 Izv. Ross. Akad. Nauk, Ser. Fiz. {\bf 64}, 541 (2000)
 [ Bull. Russ. Acad. Sci., Phys. (USA) {\bf 64}, 434 (2000) ].
%
\bibitem{SBC}
D. Sarchi, P.F. Bortignon, G. Colo, Phys. Lett. {\bf B601}, 27 (2004).
\end{thebibliography}
%

%

\end{document}